\documentclass[aps,twocolumn,superscriptaddress,showpacs]{revtex4-1}

\usepackage{amsmath,amsfonts}
\usepackage{graphicx}
\usepackage{hyperref}

\usepackage{dcolumn}

\begin{document}

\title{Topological Quantum Phase Transition in 5$d$ Transition Metal Oxide Na$_2$IrO$_3$}

\author{Choong H. Kim}
\affiliation{Department of Physics and Astronomy and Center for Strongly
  Correlated Materials Research, Seoul National University, Seoul
  151-747, Korea}

\author{Heung Sik Kim}
\affiliation{Department of Physics and Astronomy and Center for Strongly
  Correlated Materials Research, Seoul National University, Seoul
  151-747, Korea}

\author{Hogyun Jeong}
\affiliation{Computational Science and Technology Interdisciplinary Program, Seoul National University, Seoul 151-747, Korea}
\affiliation{Korea Institute of Science and Technology Information, Daejeon 305-806, Korea}

\author{Hosub Jin}
\affiliation{Department of Physics and Astronomy, Northwestern University, Evanston, Illinois 60208, USA}

\author{Jaejun Yu}
\email[Corresponding author.]{jyu@snu.ac.kr}
\affiliation{Department of Physics and Astronomy and Center for Strongly
  Correlated Materials Research, Seoul National University, Seoul 151-747, Korea}
\affiliation{Center for Theoretical Physics, Seoul National University, Seoul 151-747, Korea}

\date{\today}

\begin{abstract}
  We predict a quantum phase transition from normal to topological
  insulators in the 5$d$ transition metal oxide Na$_2$IrO$_3$, where the
  transition can be driven by the change of the long-range hopping and
  trigonal crystal field terms.  From the first-principles-derived
  tight-binding Hamiltonian we determine the phase boundary through the
  parity analysis.  In addition, our first-principles calculations for
  Na$_2$IrO$_3$ model structures show that the interlayer distance can be
  an important parameter for the existence of a three-dimensional strong
  topological insulator phase. Na$_2$IrO$_3$ is suggested to be a
  candidate material which can have both a nontrivial topology of bands and
  strong electron correlations.
\end{abstract}


\maketitle

Topological insulators are newly discovered materials with a bulk band gap
and topologically protected metallic surface states \cite{kane,bernevig}.
The theoretical predictions on Bi-based topological insulators, such as
Bi$_x$Sb$_{1-x}$, Bi$_2$Se$_3$, and Bi$_2$Te$_3$ \cite{fkm,zhang}, have
been experimentally realized \cite{koenig,hseih,xia,chen}.  The search for
topological insulators has been extended to ternary Heusler \cite{heusler}
and chalcogenide compounds \cite{chalcogenides}, but they are still
limited to either the narrow or zero gap semiconductors.  Recently the
layered honeycomb lattice Na$_2$IrO$_3$ \cite{shitade} and pyrochlore
A$_2$Ir$_2$O$_7$ \cite{pesin,yang} have been suggested as possible topological
insulators, although topological insulators with transition metal $d$
electrons have not been fully investigated.  Contrary to the $s$-$p$
electron systems, transition metal oxides with localized $d$ electrons are
expected to have both strong on-site Coulomb interactions and spin-orbit
couplings.  In particular, $5d$ transition metal oxides such as iridates
have a relatively weaker Coulomb correlation competing with spin-orbit
coupled band structures.

The interplay between the nontrivial band topology and the electron correlation effect can be an interesting development in the study of topological insulators.
Ir-based transition metal oxides have shown some noble physics,
such as the $j_{\rm eff}=1/2$ insulating state in Sr$_2$IrO$_4$ \cite{kim1,kim2,jin1},
the anomalous metal-insulator transition in A$_2$Ir$_2$O$_7$ \cite{matsuhira},
and the spin-liquid state in Na$_4$Ir$_3$O$_8$ \cite{okamoto},
where both spin-orbit coupling and correlation play important roles.

Na$_2$IrO$_3$ \cite{singh,liu} has been proposed as a layered quantum spin Hall (QSH) insulator. Assuming the $j_{\rm eff}=1/2$ character of the band near the Fermi level \cite{shitade},
a single-band tight-binding model for the Ir-O layer is mapped to the Kane-Mele model \cite{kane} for the QSH effect.
The proposed model Hamiltonian is, however, inconsistent with the first-principles band structure result
which leads to a different prediction on the band topology of Na$_2$IrO$_3$ \cite{jin2}.
The inconsistency is partly due to the structural parameters used for the first-principles calculations.
It implies that its band topology may be sensitive to the structural variation.
Consequently a small change of structure or interaction strength can drive a quantum phase transition,
e.g., the change of its topological character.
To understand the basic physics determining the band topology of Na$_2$IrO$_3$,
we need to clarify the topological character of the spin-orbit coupled ground state and its dependence on the structure and interaction strength.

In this Letter, we present a quantum phase transition between topological
insulators (TIs) and normal insulators (NIs) in Na$_2$IrO$_3$ based on an
effective tight-binding Hamiltonian.  We derived the effective Hamiltonian
from the realistic tight-binding description of first-principles band
structures.  The electronic structure of the planar edge-shared IrO$_6$
octahedra contains large trigonal crystal field and direct
hopping terms as well as a significant long-range hopping between extended Ir 5$d$ orbitals.
The phase boundary between topological and normal insulating phases is
shown to depend on both the trigonal crystal field and the long-range
hopping in Na$_2$IrO$_3$.  From the first-principles calculations of model
structures, which simulate the change of tight-binding parameters, we
confirmed that the interlayer distance can play a crucial role in the
determination of the three-dimensional strong TI in Na$_2$IrO$_3$.

\begin{figure}
  \centering
  \includegraphics[width=0.45\textwidth]{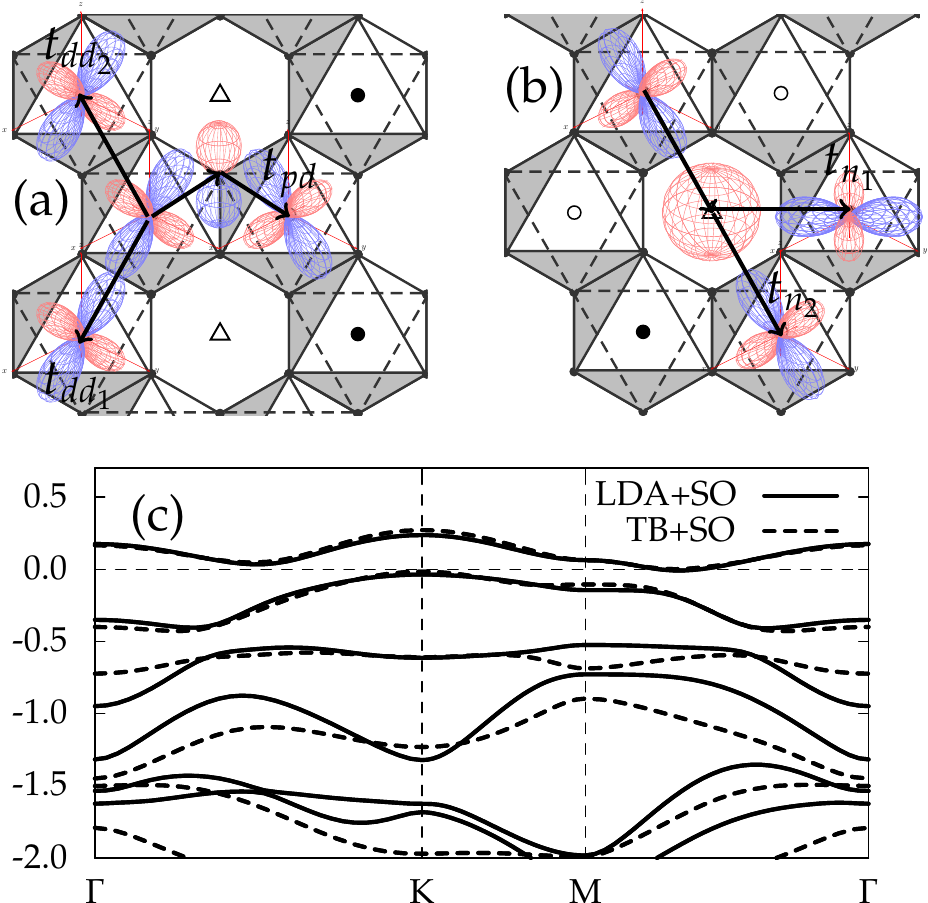}
  \caption{Hopping parameters considered in our tight-binding model :
  (a) the indirect hopping($t_{pd}$) mediated by the oxygen $2p$ orbital and
  two kinds of direct hoppings ($t_{dd_1}$ and $t_{dd_2}$),
  (b) the second-nearest-neighbor hopping ($t_{n_1}$) and the
  third-nearest-neighbor hopping ($t_{n_2}$), and
  (c) the band structure of first-principles (solid lines) and tight-binding
  (dashed lines) calculations with spin-orbit coupling.
  }
  \label{fig:tb}
\end{figure}

\begin{table}
  \centering
  \begin{ruledtabular}
  \begin{tabular}{ccD{.}{.}{2.3}}
    Parameters &  Character & \multicolumn{1}{c}{Value (eV)} \\\hline
    $t_{pd}$ &  $(pd\pi)^2/(\epsilon_d-\epsilon_p)$ & 0.25 \\
    $t_{dd_1}$ & $\frac34(dd\sigma) + \frac14(dd\delta)$ & -0.5 \\
    $t_{dd_2}$ & $\frac12(dd\pi) + \frac12(dd\delta)$ & 0.15 \\
    $t_{n}$ & $(sd\sigma)^2/(\epsilon_d-\epsilon_s)$ & -0.075 \\
    $\Delta$ & $E_{e'_{g} } - E_{a_{1g} }$ & 0.6 \\
    $\lambda$ & Spin-orbit coupling & 0.5 \\
  \end{tabular}
\end{ruledtabular}
\caption{Values of tight-binding parameters (in eV)
  obtained from the first-principles band structure.}
\label{table1}
\end{table}

We introduce a tight-binding (TB) model for Na$_2$IrO$_3$ from the results
of first-principles calculations \cite{supp}.
The model is based on the Ir $t_{2g}$
manifold in the two-dimensional honeycomb lattice and incorporates
parameters for an indirect hopping through the oxygen $2p$-orbital ($t_{pd}$),
two kinds of direct hopping between neighboring Ir atoms ($t_{dd_1}$ and
$t_{dd_2}$), and another indirect hopping through the sodium $3s$-orbital
($t_n$), as illustrated in Fig.~1(a) and (b).  It is noted that the
indirect hopping through the sodium $3s$-orbital of Fig.~\ref{fig:tb}(b)
makes the second- and third-nearest-neighbor hopping terms non-negligible
and plays a crucial role in the determination of the topology of the band
structure.  In addition, we take into account both trigonal crystal field
$\Delta$ and spin-orbit coupling (SOC) $\lambda$ terms.  The TB parameters
are summarized in Table~\ref{table1}.  The energy unit is in eV.  The
values of the 	parameters are fitted to the results of
density-functional-theory (DFT) calculations within the local density
approximation (LDA).  For the DFT calculations, we used the DFT code,
OpenMX \cite{openmx}, based on the linear-combination-of
pseudo-atomic-orbitals method \cite{ozaki}.  The SOCs were included via the
relativistic $j$-dependent pseudopotential scheme in the noncollinear DFT
formalism \cite{macdonald,bachelet,theurich}.  The unit cell we used in
the DFT calculations is a simplified version of the original crystal,
where the c-axis periodicity is reduced by changing the relative stacking
of Na layers with respect to the Ir network and also by neglecting the
distortion of oxygen octahedra.  We have checked the band structures with
the different stacking of Na layers and the rotation of oxygen atoms
around the axis perpendicular to the plane, and observed only a small
change in the band dispersion and fitting parameters.
Our TB model describes well the DFT band structure, especially the bands
near the Fermi level ($E_{\mathrm{F}}$), as shown in
Fig.~\ref{fig:tb}(c).

\begin{table}
  \centering
  \begin{ruledtabular}
  \begin{tabular}{lccccc}
    &$\delta(\Gamma)$ & $\delta(M_1)$ & $\delta(M_2)$ & $\delta(M_3)$ &$Z_2(\nu)$ \\\hline
    $t_n=-0.075$ & -1 & -1 & -1 & -1 & 0\\ 
    $t_n=0$ & -1 &  1 & 1 & 1& 1 \\
  \end{tabular}
\end{ruledtabular}
\caption{Parity invariants $\delta(\Gamma_i)$ and $Z_2$ topological invariants $(\nu)$
with and without $t_n$
as determined from the product of parity eigenvalues at each time-reversal-invariant mometum
$\Gamma$, $M_1$, $M_2$, and $M_3$.}
\label{table2}
\end{table}

It is interesting that the calculated parity invariants and $Z_2$
topological numbers for our TB model depend critically on the second- and
third-nearest-neighbor hopping term $t_n$, as demonstrated in Table
\ref{table2}.  The topological invariants were determined by following the
method proposed by Fu and Kane \cite{fukane}.  The $Z_2$ topology of the
DFT band structure with $t_n=-0.075$ turns out to be trivial.  This result
is consistent with the previous first-principles calculations result \cite{jin2},
contrary to the quantum spin Hall insulator phase predicted by Shitade {\it et al.} \cite{shitade}.
This discrepancy is likely due to an over simplification of the tight-binding model employed in Ref.~\cite{shitade}.
Their model was based on the assumption that
the low-energy degrees of freedom are determined by the half-filled $j_{\rm eff}$=1/2 doublets.
However, when the significant trigonal crystal field is introduced in Na$_2$IrO$_3$,
$j_{\rm eff}$ = 1/2 and $j_{\rm eff}$= 3/2 are not well separated and the $j_{\rm eff}$=1/2 doublets no longer serve as a useful basis \cite{jin2}.
The extended nature of the $5d$-orbital combined with
the edge-shared octahedral structure in the Ir$_{2/3}$Na$_{1/3}$O$_2$ plane of Na$_2$IrO$_3$
is a source of such a strong trigonal crystal field.

The topological character of Na$_2$IrO$_3$ is quite sensitive to the
magnitude of $t_n$.  By turning off the second- and third-nearest-neighbor
hopping, i.e., $t_n=0$, the system becomes a nontrivial TI with $\nu=1$.
The tiny difference in the magnitude of $t_n$ is responsible for the
inversion of the valance and conduction bands, thereby leading to the
change of the $\delta(M)$ sign.  It indicates that $t_n$ is a key control
parameter for the 'band-inversion' in this system.  Therefore, the TB
model with the nearest-neighbor hoppings only can not be sufficient for
the description of the topological character of Na$_2$IrO$_3$ even though
the nearest-neighbor TB models can describe reasonably the band
dispersions of the $j_{\rm eff}=1/2$ states in Sr$_2$IrO$_4$
\cite{kim1,jin1} and the hyper-Kagome Na$_4$Ir$_3$O$_8$ \cite{norman,podolsky}.

\begin{figure}
  \centering
  \includegraphics[width=0.5\textwidth]{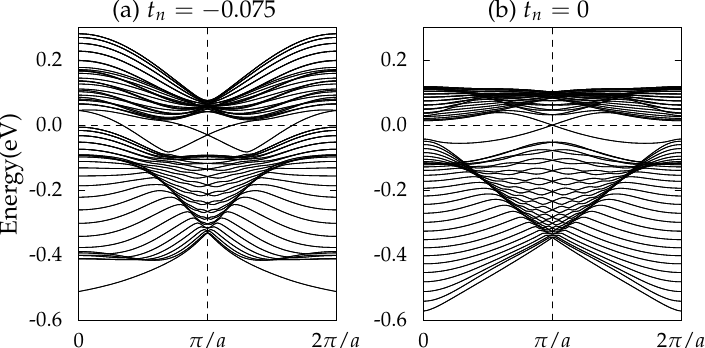}
  \caption{One-dimensional energy bands for armchair strip in the (a) NI phase
    with $t_n=-0.075$ and (b) TI phase with $t_n=0$.}
  \label{fig:edgestates}
\end{figure}

The nontrivial (trivial) $Z_2$ topological number can also be confirmed by the
odd (even) number of pairs of gapless edge states.
To examine the edge states, we constructed a TB Hamiltonian for the strip
geometry with two edges in an armchair configuration.
Figure~\ref{fig:edgestates} shows the one-dimensional energy bands with $t_n=-0.075$ and $t_n=0$.
The bulk states and gaps are clearly seen and 
there are edge states which transverse the gap.
In the normal insulating phase ($\nu=0$), the edge states cross the Fermi energy an even number of times,
as expected from the trivial $Z_2$ topological number.

\begin{figure}
  \centering
  \includegraphics[width=0.4\textwidth]{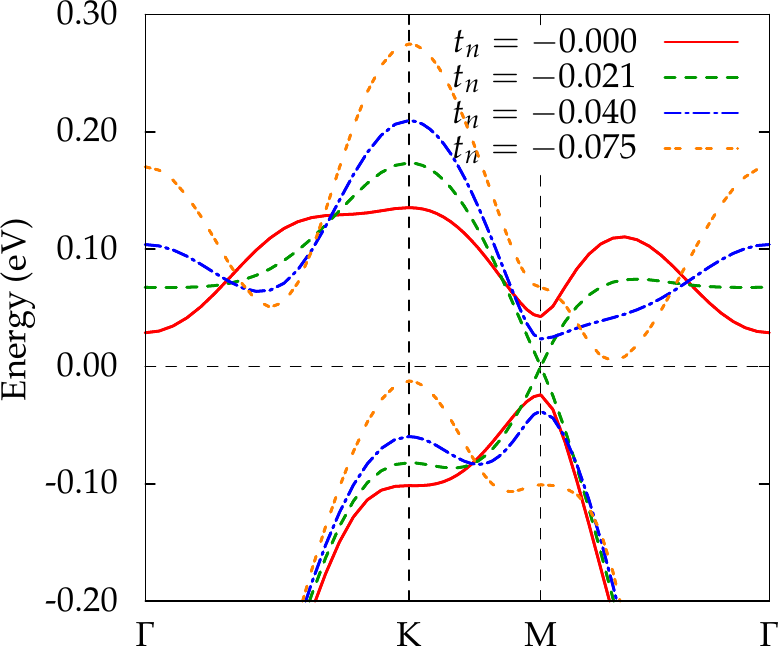}
  \caption{TB band structure with the variation of $t_n$.
    The solid (red) ($t_n=0$) lines belong to the TI,
    and the dot-dashed (blue) ($t_n=-0.040$) and double-dashed (orange) ($t_n=-0.075$) lines belong to the NI.
    Note that, at the transition point, i.e., $t_n=-0.021$, the dispersion is linear.
  }
  \label{fig:gapclosing}
\end{figure}

Transitions between trivial and nontrivial phases can be tuned by
adjusting the key parameters of the band structure. 
Murakami and co-workers \cite{murakami} developed the low-energy effective theory for
the phase transition between TI and NI systems in 2D and discussed the
classification of the possible types of transition.  Following their
classification, the gap closing at the time-reversal-invariant-momenta
($\mathbf{k}=\mathbf{G}/2$) occurs in systems with inversion symmetry.
In our model, $t_n$ controls the gap closing at the $M$ point 
in the Brillouin zone contrary to the case of a honeycomb
lattice model such as the Kane-Mele model, where the gap closes at the $K$ and $K'$ points.
The low-energy long-wavelength effective Hamiltonian, which is expanded to
linear order in $k$ around the $M$ point, can be written as
\begin{equation}
 H(\mathbf{k}) = E_0 +
\begin{pmatrix}
m& z_{-} & & \cr
z_{+} & -m & & \cr
 & & m & -z_{+}\cr
 & & -z_{-} & -m \cr
\end{pmatrix},
\end{equation}
where $\textbf{k}'=\mathbf{k}-M$ and $z_\pm=b_1 k'_x + b_3 k'_y \pm i
b_2k'_y$ with real constant $b_1$, $b_2$ and $b_3$ \cite{murakami}.  Since
the eigenenergies $E=E_0\pm\sqrt{m^2+z_{+}z_{-}}$, $m=0$ corresponds to
the gap closing.

Figure \ref{fig:gapclosing} shows the TB band structure along some
high-symmetry line for the several values of $t_n$.  In the TI region
($t_n=0$), the valance and conduction bands are separated by a finite gap
with negative mass.  When $|t_n|$ increases, the direct gap at the $M$
point decreases.  Finally, the gap collapses at the transition point
$t_n\simeq-0.021$.  As $|t_n|$ passes through the transition point, the
gap reopens and the system becomes a normal insulator with positive mass.
During this procedure, the order of the bands is inverted around the $M$
point, which characterizes the topological nature of the system.

To further verify the relation between the mass and the control parameters,
we obtain the expression for the mass as a function of $t_n$, $\Delta$ and SOC ($\lambda$):
\begin{equation}\label{eq:mass}
2m = A + B \Delta+ C t_{n_1} - 2t_{n_2} + D \Delta t_{n_1}.
\end{equation}
Here, to clarify the role of second- and third-nearest hoppings, we refined
$t_n$ into the second-nearest-neighbor hopping, $t_{n_1}$, and the third-nearest-neighbor hopping, $t_{n_2}$.
The expansion coefficients $A$, $B$, $C$, and $D$ can be represented as functions of $\lambda$.
In the range of $\lambda$ from $0.8\lambda_0$ to $1.6\lambda_0$, where
$\lambda_0$ is the SOC determined by the LDA calculation, the coefficients
can be fitted as follows:
\begin{align}
A &= +0.0746 - 0.1438 \lambda\nonumber\\
B &= -0.0237 - 0.3061 \lambda + 0.2587 \lambda^2\nonumber\\
C &= -2.2178 + 1.2660 \lambda\nonumber\\
D &= +1.0975 - 0.6837\lambda.
\end{align}
In the limiting cases of $\lambda/\lambda_0=1$ and $\infty$, the coefficients are also shown
in Table~\ref{table3}.  From Eq.~(\ref{eq:mass}), we can determine the
$Z_2$ topological number of the system for given $t_n$, $\Delta$, and
$\lambda$ by checking the sign of mass.  This mass function correctly
characterizes the topological phase transition.

\begin{table}
  \centering
  \begin{ruledtabular}
    \begin{tabular}{ccccc}
      $\lambda/\lambda_0$ & $A$ & $B$ & $C$ & $D$\\\hline
      1 & -0.00 & -0.12 & -1.61 & 0.58\\
      $\infty$ & -0.13 &-0.00 & -0.13 & 0.00 
    \end{tabular}
  \end{ruledtabular}
\caption{The expansion coefficients of mass function in Eq.~\ref{eq:mass} when $\lambda/\lambda_0=0$ and $\infty$.
  These coefficients at $\lambda/\lambda_0=1$ and $\Delta=0.6$ correctly describe the transition point $t_n\simeq-0.021$.
}
\label{table3}
\end{table}

\begin{figure}
  \centering
  \includegraphics[width=0.45\textwidth]{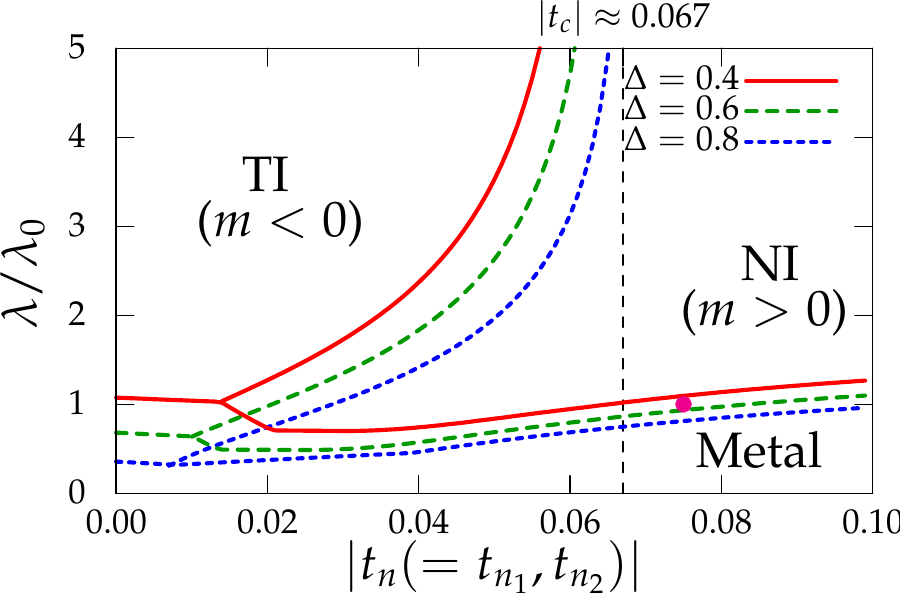}
  \caption{Phase diagram as a function of $t_n$ and SOC for several values of $\Delta$.
    The pink circle indicates where the reality exists.
    $|t_c|=0.062$ is the asymptotic line in the case of $\Delta=0.6$.
  }
  \label{fig:diagram}
\end{figure}

Figure~\ref{fig:diagram} shows a $\lambda-t_n$ phase diagram for several
values of $\Delta$ .  For a sufficiently large $\lambda$, an indirect bulk
gap opens and the system becomes an insulator.  The insulating region is
divided into two phases: TI and NI.  The phase boundary between TI and NI
corresponds to the $m=0$ line.  There seems to be an asymptotic limit of
$t_n$ beyond which the system stays as a normal insulator regardless of
the strength of the SOC $\lambda$.  The vertical dashed line in
Fig.~\ref{fig:diagram} is described by
\begin{equation}
t_n = \frac{A}{C-2} \approx -0.067
\end{equation}
with $\Delta$=0.6.  This implies that an arbitrarily large SOC cannot
guarantee the nontrivial topology even if the upper four bands originate
from the $j_{\rm eff}=1/2$ states.

Another important factor controlling the topological character is the
trigonal crystal field, $\Delta$.  In fact, both SOC and the trigonal crystal
field are two competing parameters characterizing the bands near the Fermi
level.  While SOC prefers the formation of the $j_{\rm eff}=1/2$ state,
the large trigonal crystal field becomes an obstacle for the $j_{\rm
  eff}=1/2$ state.  In this system, however, the trigonal crystal field
seems to play a crucial role.  As far as $B<0$ and $D>0$ of
Eq. (\ref{eq:mass}) as shown in Table \ref{table3}, it is clear that a
large trigonal crystal field is in favor of the nontrivial topology. This
trend in the parameter dependence of the topological character is clearly
reflected in the phase diagram of Fig.~\ref{fig:diagram}.

Returning to a realistic system of Na$_2$IrO$_3$ materials, we tried to
probe possible TI phases by performing first-principles calculations.
According to the simulated TB parameters for the TI phase, we choose two
representative structures among the various structures: (i) the original
geometry of Na$_2$IrO$_3$ ($c/c_0=1.0$) and (ii) a virtual structure with
the interlayer distance expanded by 30\% ($c/c_0=1.3$).  Calculated $Z_2$
topological numbers for each configuration are as listed in
Table~\ref{table4}.  Here the $c/c_0$ parameter represents the change of
interlayer distance relative to the original one.  It is remarkable to
find a nontrivial $Z_2$ number for the structure with $c/c_0=1.3$.  In
other words, the increased interlayer distance drives the NI into the TI.
To understand the change of electronic structure between $c/c_0=1.0$ and
$c/c_0=1.3$, we constructed the maximally localized Wannier function
(MLWF) \cite{marzari,weng} and obtained the MLWF effective Hamiltonian.
It is found that the trigonal crystal field enhanced by the increase of
interlayer distance contributes to the change of $Z_2$ character for
Na$_2$IrO$_3$.  When we increase the interlayer distance, the energy
level of the interplane Na $3s$ state lowers.  The second- or higher-order
hoppings through the unoccupied $3s$ state of interplane Na contribute to
the on-site Hamiltonian matrix elements.  Since these matrix elements give
rise to the energy separation between $e'_{g}$ and $a_{1g}$ states, it can
be interpreted that the trigonal crystal field contributes to the band
inversion at the $M$ point.
In fact the band inversion occurs only at three $M$ points in the $k_z=0$ plane.
However, as shown in Table~\ref{table4},
the parity invariants $\delta(\Gamma_i)$ for $c/c_0=1.3$ are quite different
for the $k_z=0$ and $k_z=\pi$ planes.  
It indicates that the $c$-axis hopping is also important for the existence of the
TI phase in the three-dimensional system of Na$_2$IrO$_3$.

\begin{table}
  \centering
  \begin{ruledtabular}
  \begin{tabular}{lccccc}
    &$\delta(\Gamma)$ & $\delta(M)$ & $\delta(A)$ & $\delta(L)$ &$\nu;(\nu_1\nu_2\nu_3)$ \\\hline
    $c/c_0$=1.0 & -1 & -1 & -1 & -1 & 0; (000)\\ 
    $c/c_0$=1.3 & -1 &  1 & -1 & -1 & 1; (000) \\
  \end{tabular}
\end{ruledtabular}
\caption{Parity invariants $\delta(\Gamma_i)$ and $Z_2$ topological invariants calculated from first-principles calculations.
We considered two different structures: (i) the original geometry of
Na$_2$IrO$_3$ ($c/c_0=1.0$) and (ii) a virtual structure with the interlayer distance enlarged by 30\% ($c/c_0=1.3$).
}
\label{table4}
\end{table}

In conclusion we provide a microscopic picture for the topological phase
diagram in Na$_2$IrO$_3$ and identify the key control parameters based
on the effective Hamiltonian analysis.  We predict that the TI phase of
Na$_2$IrO$_3$ can be realized by controlling the long-range hopping and
trigonal crystal field terms.  Our first-principles calculations for the
simulated Na$_2$IrO$_3$ model structure suggest that the interlayer
distance can play a crucial role in the determination of a three-dimensional
strong TI.  In practice, we propose two ways of driving a transition from NI to TI:
(i) epitaxial strain by the substitution of Na by other elements such as Li or a film growing on a appropriate substrate or
(ii) intercalation of some molecules for the increase of the interlayer distance
(e.g., is the water-intercalated Na$_{0.35}$CoO$_2$$\cdot$1.3H$_2$O showing the superconductivity \cite{takada}).
Indeed our first-principles calculations demonstrate that
the topological insulator phase of Li$_2$IrO$_3$ can be achieved by 2$\%$ in-plane lattice strain \cite{hkim}.
In addition, other structure manipulation techniques
developed for superlattice and heterostructure may be adopted to control
the NI-to-TI transition and to design the topological-insulator-based devices.
A recent experiment observed an antiferromagnetic (AFM) insulating behavior
in Na$_2$IrO$_3$ \cite{singh,liu}.
The AFM ordering in the TI phase of Na$_2$IrO$_3$ can be a candidate of a topological magnetic insulator
\cite{li,wang} or a topological Weyl semimetal \cite{wan} below ordering temperature ($T_N$).
Above $T_N$, correlations could enhance the
SOC effects due to the suppression of effective bandwidth to stabilize the TI phase \cite{pesin}.
Further there are proposals for the topological Mott insulator having gapless surface spin-only excitations
\cite{shitade,pesin}.
Na$_2$IrO$_3$ is likely to provide a new playground to study the effect of the correlation in TI.


\begin{acknowledgments}
This work was supported by the NRF through the ARP (R17-2008-033-01000-0).
We also acknowledge the support from KISTI under the Supercomputing Application Support Program.
\end{acknowledgments}

\end{document}